


\font\bigbold=cmbx12
\font\eightrm=cmr8
\font\sixrm=cmr6
\font\fiverm=cmr5
\font\eightbf=cmbx8
\font\sixbf=cmbx6
\font\fivebf=cmbx5
\font\eighti=cmmi8  \skewchar\eighti='177
\font\sixi=cmmi6    \skewchar\sixi='177
\font\fivei=cmmi5
\font\eightsy=cmsy8 \skewchar\eightsy='60
\font\sixsy=cmsy6   \skewchar\sixsy='60
\font\fivesy=cmsy5
\font\eightit=cmti8
\font\eightsl=cmsl8
\font\eighttt=cmtt8
\font\tenfrak=eufm10
\font\sevenfrak=eufm7
\font\fivefrak=eufm5
\font\tenbb=msbm10
\font\sevenbb=msbm7
\font\fivebb=msbm5
\font\tensmc=cmcsc10


\newfam\bbfam
\textfont\bbfam=\tenbb
\scriptfont\bbfam=\sevenbb
\scriptscriptfont\bbfam=\fivebb

\newfam\frakfam
\textfont\frakfam=\tenfrak
\scriptfont\frakfam=\sevenfrak
\scriptscriptfont\frakfam=\fivefrak


\def\eightpoint{%
\textfont0=\eightrm   \scriptfont0=\sixrm
\scriptscriptfont0=\fiverm  \def\rm{\fam0\eightrm}%
\textfont1=\eighti   \scriptfont1=\sixi
\scriptscriptfont1=\fivei  \def\oldstyle{\fam1\eighti}%
\textfont2=\eightsy   \scriptfont2=\sixsy
\scriptscriptfont2=\fivesy
\textfont\itfam=\eightit  \def\it{\fam\itfam\eightit}%
\textfont\slfam=\eightsl  \def\sl{\fam\slfam\eightsl}%
\textfont\ttfam=\eighttt  \def\tt{\fam\ttfam\eighttt}%
\textfont\bffam=\eightbf   \scriptfont\bffam=\sixbf
\scriptscriptfont\bffam=\fivebf  \def\bf{\fam\bffam\eightbf}%
\abovedisplayskip=9pt plus 2pt minus 6pt
\belowdisplayskip=\abovedisplayskip
\abovedisplayshortskip=0pt plus 2pt
\belowdisplayshortskip=5pt plus2pt minus 3pt
\smallskipamount=2pt plus 1pt minus 1pt
\medskipamount=4pt plus 2pt minus 2pt
\bigskipamount=9pt plus4pt minus 4pt
\setbox\strutbox=\hbox{\vrule height 7pt depth 2pt width 0pt}%
\normalbaselineskip=9pt \normalbaselines
\rm}


\def\pagewidth#1{\hsize= #1}
\def\pageheight#1{\vsize= #1}
\def\hcorrection#1{\advance\hoffset by #1}
\def\vcorrection#1{\advance\voffset by #1}

\newcount\notenumber  \notenumber=1              
\newif\iftitlepage   \titlepagetrue              
\newtoks\titlepagefoot     \titlepagefoot={\hfil}
\newtoks\otherpagesfoot    \otherpagesfoot={\hfil\tenrm\folio\hfil}
\footline={\iftitlepage\the\titlepagefoot\global\titlepagefalse
           \else\the\otherpagesfoot\fi}

\def\abstract#1{{\parindent=30pt\narrower\noindent\eightpoint\openup
2pt #1\par}}
\def\smc{\tensmc}


\def\note#1{\unskip\footnote{$^{\the\notenumber}$}
{\eightpoint\openup 1pt
#1}\global\advance\notenumber by 1}

\def\frac#1#2{{#1\over#2}}
\def\dfrac#1#2{{\displaystyle{#1\over#2}}}
\def\tfrac#1#2{{\textstyle{#1\over#2}}}
\def\({\left(}
\def\){\right)}
\def\<{\langle}
\def\>{\rangle}
\def\2pd#1#2#3{\frac{\partial^2#1}{\partial#2\partial#3}}

\def\sqr#1#2{{\vcenter{\vbox{\hrule height.#2pt
        \hbox{\vrule width.#2pt height#1pt \kern#1pt
           \vrule width.#2pt}
        \hrule height.#2pt}}}}
\def\square{\mathchoice\sqr64\sqr64\sqr{4.2}3\sqr33}
\def\ni{\noindent}
\def\ref #1{$^{[#1]}$}
\def\slash{\!\!\!\!/}
\def\lqq{\lq\lq}
\def\rqq{\rq\rq}
\def\dperp{\delta^\perp}

\def\a{\alpha}

\def\d{\delta}
\def\dperp{\d^\perp}

\def\phys{{\hbox{\sevenrm phys}}}
\def\anti{{\hbox{\sevenrm anti}}}
\def\L{{\cal L}}
\def\N{{\cal N}}

\def\ket#1{|#1\>}


\pageheight{24cm}
\pagewidth{15.5cm}
\hcorrection{-2.5mm}
\magnification \magstep1
\baselineskip=16pt
\parskip=5pt plus 1pt minus 1pt
%
%
\rightline {MZ-TH/93-02}
\rightline {DIAS-STP-93-03}
\rightline {Revised Version}
\vskip 40pt
\centerline{\bigbold A NEW SYMMETRY FOR QED}
\vskip 30pt
\centerline{\smc Martin Lavelle{\hbox {$^*$}}{\note{e-mail:
lavelle@vipmza.physik.uni-mainz.de}}
and  David McMullan{\hbox {$^{\dag}$}}{\note{e-mail: mcmullan@stp.dias.ie}}}
\vskip 15pt
{\baselineskip 12pt \centerline{\null$^*$Institut f\"ur Physik}
\centerline{Johannes Gutenberg-Universit\"at}
\centerline{Staudingerweg 7, Postfach 3980}
\centerline{W-6500 Mainz, F.R.\thinspace Germany}
\vskip 12pt
\centerline{\null$^{\dag}$Dublin Institute for Advanced Studies}
\centerline{School of Theoretical Physics}
\centerline{10 Burlington Road}
\centerline{Dublin 4}
\centerline{Ireland}
}
\vskip 7truemm
\vskip 40pt
{\parindent=0.5in\narrower\ni{\bf Abstract}\quad We demonstrate that
QED exhibits a previously unobserved symmetry. Some consequences
are discussed.
\bigskip\bigskip
\ni{\bf PACS No.'s:}\quad 12.20.--m, 11.30.--j
\par}
\bigskip
\centerline{Submitted to Physical Review Letters}
\vfill\eject
\noindent Quantum Electrodynamics (QED) is the cornerstone of modern high
energy physics. A generalisation of its gauge invariance is found in all
other theories of nature. In this letter we shall show that QED displays a
further, quite distinct, symmetry.

In order to quantise QED  gauge fixing is essential. Working in Feynman
gauge (as we shall throughout this letter) the Lagrangian is
$$
\L=-\tfrac14 F_{\mu\nu}F^{\mu\nu} - \tfrac12 (\partial_\mu A^\mu)^2
+\bar\psi(i D\slash-m)\psi + i\bar c \,\square\,   c\,,
\eqno (1)
$$
where $D_\mu=\partial_\mu+ i g A_\mu$ and the ghosts are Hermitian.
Although the ghost fields decouple, they are retained in (1) since it
is this formulation of QED that can be extended to nonabelian
theories rather than, say, the ghost free Gupta-Bleuler description.

As is well known the photon has two, transverse, degrees of freedom.
However, in the covariant formulation (1) all four components of the gauge
field are present. One can argue that the gauge fixing term has accounted
for one degree of freedom, while the ghosts heuristically contribute one
negative degree of freedom. Thus the full Lagrangian (1) does indeed
describe the interaction of fermions with the two degrees of freedom of the
photon. This line of argument can be put on a firmer footing by exploiting
the BRST invariance of the QED Lagrangian.

The BRST transformations are\ref{1}
$$
\eqalign{\d A_\mu & =  \partial_\mu c\,,\cr
\d c & =  0\,,\cr
\d\bar c& =  -i\partial_\mu A^\mu\,,\cr
\d\psi& =-i g c\psi\,,\cr
\d\bar\psi&=-i g \bar\psi c\,,
}\eqno (2)
$$
and we note that the BRST transform of a field has increased its
ghost number by one.
The invariance of (1) under these transformations can be shown in three steps:
the original photonic and Dirac
Lagrangians are each separately invariant (reflecting the close
connection between gauge invariance and BRST invariance), while the gauge
fixing and ghost terms together are invariant.
One of the most remarkable properties of the BRST transformation is that it
vanishes when applied twice to any field; $\d^2\equiv 0$. This nilpotency
property can be
seen from (2) using the (uncoupled) ghost equation of motion. (This reliance on
the  equations of motion can be removed by introducing auxiliary fields as
in Ref.\thinspace 1.)

The BRST invariance of the Lagrangian (1) allows for a succinct
characterization of the physical states of the theory. The
transformations (2) are generated by the conserved, Hermitian charge
$$
Q:=\int d^3x\,\bigl(-\dot\pi_0(x)c(x)+\pi_0(x)\dot{\bar
c}(x)\bigr)\,,\eqno(3)
$$
where $\pi_0:=-\partial_\mu A^\mu$. The
physical states are then identified with those states, $\ket\psi$,
which satisfy $Q\ket\psi=0$. This is not the whole story,
though, since the nilpotency of the BRST transformation implies that
$Q^2=0$, hence any state of the form $\ket\psi=Q\ket\chi$, for any
$\ket\chi$, will trivially be physical in this sense.
More properly, then, the physical states should be
identified with the quotient of the BRST invariant (closed) states
with these trivial (exact) states.  Working on the Fock space
for this theory constructed out of the free fields (which are
identified with the asymptotic in and out states), it was shown in
Ref.\thinspace1 that the photonic sector of the physical states
could be
identified with the states built from the transverse components of the photon.
Even in this abelian theory this is quite an involved analysis.

This Fock space discussion is only strictly relevant to the free
theory: when matter is present the infrared structure of the
theory\ref{2}
implies that asymptotic states which have an
arbitrary number of photons in them must be allowed. The direct extension
of the arguments presented in Ref.\thinspace1 to such coherent states is
then not clear.

This use of the BRST charge to characterize the physical states is
familiar in mathematics and would be called a cohomology
theory---the physical states then being the zeroth cohomology of the
appropriate complex. Thus the BRST charge is playing a role similar
to the exterior derivative, $d$, acting on differential forms.
Mathematicians have developed many techniques for analysing such
structures (see, for example, Ref.\thinspace3), the most powerful of
which is to introduce an adjoint operation (denoted by $d^*$ for
the differential forms example). This is also nilpotent and can be
used to refine the description of the $d$-closed forms.
It would be useful to have a similar adjoint to the BRST
charge, however, it is not clear that such an object
can be constructed since, in addition to its various algebraic
properties, it must also be conserved if it is to have any physical
significance.

The anti-BRST transformation\ref{1}, $\d_\anti$, is an example of a type of
adjoint to the BRST transformation. Its existence relies on the
simple fact that the Lagrangian (1) is invariant under the
interchange $c\to i\bar c$ and $\bar c\to ic$. Acting on the
fields $\d_\anti$ essentially reproduces (2) but with the ghost and
anti-ghost interchanged. Clearly this transformation has the
property that it reduces the ghost number
of a state by one. However, its close connection with the BRST
charge, and hence the gauge invariance of the theory, means that
it has few of the useful properties desired from an
adjoint. In particular, it anticommutes with the BRST charge,
$\d\d_\anti+\d_\anti\d=0$, thus there is no analogue of the
Laplacian $dd^*+d^*d$ and the related harmonic description of cohomology
found in differential geometry.

In Ref.\thinspace4 it was argued that, for simple quantum mechanical
systems, the appropriate concept of an adjoint (or more properly, a
dual) to the BRST transformation
is not one base on the existence of pairing betweed states (as in
the relationship between $d$ and $d^*$), but
rather it is a transformation that is
compatible with the gauge fixing conditions. An extension
of that argument  to QED would suggest that we are looking for a symmetry
transformation $\dperp$ of (1) such that, as well as decreasing
the ghost number of the fields by one,
acting on the gauge fixing condition it gives zero:
$$
\dperp(\partial_\mu A^\mu)\equiv0\,.\eqno(4)
$$
(In a formulation with the auxillary field $\pi_0$, this condition
could be replaced with the weaker condition that $\pi_0+\partial_\mu
A^\mu$ is invariant.)
We say that relation (4) is dual to the BRST invariance of gauge
invariant quantities, i.e, dual to the relations $\d\bigl(\frac14
F^{\mu\nu}F_{\mu\nu}\bigr)=0$  and
$\d\bigl(\bar\psi(iD\slash-m)\psi\bigr)=0$.

The simplest way to ensure (4) is to take $\dperp A_0=i\bar c$ and
$\dperp A_i=i\dfrac{\partial_i\partial_0}{\nabla^2}\bar c$, where
$\nabla^2=\partial_i\partial_i$. So we see that solving (4) forces
us to abandon convariance and locality. The lack of covariance in
this transformation should not come as too much of a surprise
since we know that
the two transverse physical photonic states cannot be expressed covariantly.
Extending this transformation to the other fields is far from unique,
however, requiring that we have a symmetry of (1) allows
us to finally arrive at the set of transformations:
$$
\eqalign{\dperp A_0 & = i \bar c\,,\cr
\dperp A_i& =i \frac{\partial_i\partial_0}{\nabla^2}\bar c\,,\cr
\dperp c& =   A_0-\frac{\partial_i\partial_0}{\nabla^2} A_i +
\frac{g}{\nabla^2}J_0\,,\cr
\dperp \bar c& = 0\,,\cr
\dperp\psi&=  \(\frac{g}{\nabla^2}\partial_0 \bar c\) \psi\,,\cr
\dperp\bar\psi&=  \bar\psi\frac{g}{\nabla^2}\partial_0 \bar c\,,}
\eqno (5)
$$
where $J_0$ is the current density $\bar\psi\gamma_0\psi$.
Just as was the case for the BRST transformations this transformation
can be seen to be nilpotent, i.e., $(\dperp)^2\equiv0$, with the aid of the
anti-ghost equation of
motion. (Again this dependence may be avoided by introducing an auxiliary
field.)

It is important to clearly spell out why this is a symmetry of the
Lagrangian (1).
By construction, it is
now the gauge fixing term that is invariant while the photonic and
Dirac parts of the Lagrangian need, in addition, the ghost term to be
invariant.
Under the transformation (5) the Lagrangian (1) changes by a total
divergence
$$
\dperp \L=\partial_\mu \Lambda^\mu\,,
\eqno (6)
$$
where
$$\openup1\jot
\eqalign{
\Lambda^0&=i\partial^0\bar c\,\dperp c-i\bar c\,\partial^0\dperp c\,,\cr
\Lambda^i&=i\partial^i\bar c\,\dperp c-i\bar c\,\partial^i\dperp c
-i\,\square\,\bar c\,(\frac{\partial_0
A^i}{\nabla^2}-\frac{\partial^iA_0}{\nabla^2})\,.
}
\eqno (7)
$$
Note that,  due to the presence of the non-local,
$\frac1{\nabla^2}$,
terms in $\Lambda^\mu$, we cannot immediately deduce from (6) that
this is a symmetry. This is because, in general, any transformation of
a Lagrangian can be written as a total divergence if we allow non-local
terms, i.e., we may always write an arbitrary function $F$ as
$F=\partial_\mu(\dfrac{\partial^\mu}{\square} F)$.
The fundamental quantity in the quantum theory is the action, thus
we must now check that our transformation indeed preserves
the action.

Recall that we want to construct the action  from the
Lagrangian (1) where
$$
S=\int_{T_1}^{T_2}\!\! dt\!\!\int_{B_r}\!\! d^3x\,\L\,,
\eqno (8)
$$
and we are calculating the action between two times $[T_1,T_2]$
and,
for the moment, we have a spatial ball, $B_r$, of radius $r$. We will want to
take the $r\to\infty$ limit, but we will not need to take the
$[T_1,T_2]\to[-\infty,\infty]$ limit.
Under the change (5) we get
$$
S\to
S+\Lambda(T_2)-\Lambda(T_1)+ \int_{T_1}^{T_2}\!\! dt\!\!
\int_{B_r}\!\!d^3x\,\partial_i\Lambda^i\,,
\eqno (9)
$$
where
$$
\Lambda(T)=\int_{B_r}\!\!d^3x\Lambda^0(T,x)\,.
\eqno (10)
$$
The $\Lambda(T)$ terms now depend on the whole of the space slices
at the end points in time; however, this will clearly not affect the
dynamics between
the initial and final times. The
last term is potentially dangerous since it will alter the action
between the end points, and thus could alter the dynamics.
We use Stokes theorem to write
$$
\int_{B_r}\!\!d^3x\,\partial_i\Lambda^i=\int_{S^2_r}\!\!d^2\sigma_i
\Lambda^i\,.
\eqno (11)
$$
If the
$\Lambda^i$'s were local this would tend to zero as $r\to \infty$
since the fields fall off to
zero at infinity. But since the $\Lambda^i$'s are non-local, then even as
$r\to\infty$, they will receive contributions
from the whole of the spatial slice --- so (11) might not vanish.
However, we are dealing with terms of the form
$$
\int_{S_r^2}\!\!d\sigma^i\frac{f(x)}{\nabla^2}=-\frac1{4\pi}\int_{S^2_r}\!\!d\sigma
^i\!\!\int d^3y\frac{f(y)}{|x-y|}\,.
\eqno (12)
$$
Now as $r\to\infty$, i.e., as $x\to\infty$, for finite $y$ we have
$\dfrac1{|x-y|}\to0$. This is not the case, though, if $y$ and $x$ are
close.
But then, for large $x$, $f(y)\to0$ since we demand good boundary
conditions on the
fields.  We conclude that even though our symmetry is non-local, it is
indeed a symmetry in the non-trivial sense.

Before briefly discussing some of the consequences of this new
symmetry, we note that its action on the gauge fields and fermions
can formally be derived from the BRST transformations (2) under the
substitution $c\to\dfrac{i\bar c}{\partial_0}$ and by using the
equations of motion. The action on the ghosts then follow from the
required invariance of the Lagrangian. It is not clear to us if
this trick sheds any
light on this new symmetry. Indeed the use of the classical equations of motion
in a transformation does not guarantee the invariance of the action.
For example, if the classical equations of motion are used then we
can also replace $\dperp c$ in (5) by
$$
\dperp_\a c=A_0-\a\frac{\partial_i\partial_0}{\nabla^2} A_i +
\frac{g}{\square}\(\a\frac{\partial_0\partial_0}{\nabla^2}-1\)J_0\,.
\eqno(13)
$$
For all $\a$ this transformation will preserve the classical
equations of motion. However, the Lagrangian only transforms into a
total divergence {\it without use of the classical equations of
motion} for the specific choice $\a=1$.
\bigskip
We now wish to analyse some consequences of this symmetry. Given
that this is a symmetry we can use it, in much the same way as BRST
is used, to generate Ward type identities. So consider the identity
$$
<T\bigl(A_0(x)c(y)\bigr)>\equiv0\,.
\eqno (14)
$$
Applying $\dperp$ to this yields
$$
-iD(q)+D_{00}(q)-\frac{q_iq_0}{\vec q^{\,2}}D_{i0}(q)-\frac{g}{\vec
q^{\,2}}\Lambda_{00}(q)=0\,,
\eqno(15)
$$
where
$$\eqalign{
D(q)&:=\int d^4x\,e^{iq{\cdot}x}\<T\bigl(c(x)\bar c(0)\bigr)\>\,,\cr
D_{\mu\nu}(q)&:=\int d^4x\,e^{iq{\cdot}x}\<T\bigl(A_\mu(x)A_\nu(0)\bigr)\>
\,,\cr
\Lambda_{00}(q)&:=\int
d^4x\,e^{iq{\cdot}x}\<T\bigl(A_0(x)J_0(0)\bigr)\>\,.
}\eqno(16)
$$
This is not one of the usual covariant Ward identities.
However, it is straightforward to check that
it is indeed
fulfilled in QED in the Feynman gauge. The non-covariance of this
identity is a consequence of the non-covariance of the symmetry and
of the non-covariance of the physical fields.
A fuller account of the identities derivable from this new symmetry
will be presented elswhere.

Returning now to our original motivation for searching for
this symmetry, we recall that physical
states have been characterized as being BRST invariant. To refine
this description we
now impose an additional condition that physical fields must also be
invariant under the symmetry (5). Thus the physical states of the
theory must satisfy the conditions
$$
Q\ket\psi=Q^\perp\ket\psi=0\,,\eqno(17)
$$
where the anti-Hermitian charge $Q^\perp$, which generates the
transformations (5),
is given by
$$
Q^\perp:=\int d^3x\,\bigl(-i\frac{\dot\pi_0(x)}{\nabla^2}\dot{\bar
c}(x)+i\pi_0(x)\bar c(x)\bigr)\,,\eqno(18)
$$
Combining these requirements we see that a physical
state will, in addition, satisfy
$$
\N \ket\phys = 0\,,
\eqno (19)
$$
where the Hermitian Laplacian type operator $\N$ is given by
$$
\N:=QQ^\perp + Q^\perp Q\,.
\eqno (20)
$$
Although this does not have all the properties one might wish from a
Laplacian, in that $\N c=\N\bar c=0$, it does imposes additional
restrictions on the photonic states and
one can show\ref{5} that in the coherent space approach, the
physical states
are built up from the transverse components of the photon and the
fields
$$
\eqalign{
\psi_\phys(x)&=\exp\(-ig\frac{\partial_i A^i(x)}{\nabla^2}
\)\psi(x)\,,\cr
\bar\psi_\phys(x)&= \exp\( ig\frac{\partial_i A^i(x)}{\nabla^2} \)
\bar\psi(x) \,,
}\eqno (21)
$$
which are Dirac's physical electrons\ref{6}.
\bigskip
In summary we have found a new symmetry for QED. This symmetry is
non-covariant and non-local. Evidently just as BRST has a partner
in anti-BRST symmetry, where ghosts and anti-ghosts are interchanged,
an \lq anti-version\rq\ of this symmetry is easily constructed.
The $\dperp$-symmetry  may be used to refine the
characterization of physical states given by the BRST charge.
We have also shown that this new symmetry will
generate new Ward type identities in the quantum theory.
The extension of this symmetry to other gauges,
including non-covariant ones, and the geometric role of this symmetry will
be presented elsewhere.
\vfill
\eject
\ni{\bf Acknowledgements} MJL thanks the Dublin Institute for Advanced
Studies for their hospitality and the Graduierten Kolleg of Mainz
University for support. DM thanks the University of Mainz for hospitality.
We both wish to thank the first referee for pointing out an error in
an earier version of this letter.
\bigskip
\ni{\bf References}
\item{1)}{N. Nakanishi and I. Ojima, \lqq Covariant Operator Formalism
of Gauge Theories and Quantum Gravity\rqq, (World Scientific, Singapore
1990).}
\item{2)}{T.W.B. Kibble, in \lqq Mathematical Methods in Theoretical
Physics\rqq, ed.s K.T. Mahanthappa and W.E. Brittin (Gordon and
Breach, New York, 1969).}
\item{3)}{R. Bott and L.W. Tu, \lqq Differential Forms in Algebraic
Topology\rqq, (Springer-Verlag, New York, 1982).}
\item{4)}{D. McMullan, Commun. Math. Phys. 149, (1992) 161.}
\item{5)}{M. Lavelle and D. McMullan, \lqq Physical States and
Greens Functions in Gauge Theories: I QED\rqq, in preparation.}
\item{6)}P.A.M. Dirac, \lqq Principles of Quantum Mechanics\rqq, (OUP,
Oxford 1958), page 302.
\bye